\documentclass{elsart1p}

 \usepackage{graphicx}

\usepackage{amssymb}

\begin{document}

\begin{frontmatter}

\title{Interplay between Yukawa and Tomonaga\\
 in the Birth of Mesons}


\author{Toshimitsu Yamazaki}

\address{Nishina Memorial Foundation, Honkomagome 2-28-45, Bunkyo-ku, Tokyo 113-8941, Japan\\
Department of Physics, University of Tokyo, Bunkyo-ku, Tokyo 113-0033, Japan\\
RIKEN Nishina Center, Wako-shi, Saitama-ken 351-0198, Japan}

\begin{abstract}

Light is shed on the early stage in the birth of Yukawa's meson theory, particularly on the interplay between Yukawa and Tomonaga in 1933. The discovery of the muon by Nishina' group in 1937 is also  reviewed. It is pointed out that Heisenberg's attempt to explain the nuclear force in terms of the Heitler-London scheme, overcome by Yukawa and abandoned since then, is now being revived as a mechanism for a super strong nuclear force caused by a migrating real $\bar{K}$ meson.  
\end{abstract}

\end{frontmatter}

\section{Introduction}
 
It is a great honor and pleasure to have an opportunity to give a talk in 
this Yukawa Centennial Session. Here, I concentrate on the early stage of the birth of the Yukawa meson. As explained in the previous talk by Professor Sato, Yukawa and Tomonaga were physics classmates at Kyoto University. Very recently, private communications between them in 1933, the eve of the birth of Yukawa's  theory, were found \cite{Nishina-Letters}. These tell us about a very interesting and intriguing interplay and interaction between these two great physisists, then only students. It is the purpose of my talk to convey my own excitement and impressions from their premature years.

\section{The Year 1933 of Advent}

Right after the discovery of the neutron by Chadwick in 1932, Heisenberg published his famous work on the nuclear force and nuclear binding phenomena in 1932 \cite{Heisenberg:32}. This work affected young Yukawa and Tomonaga greatly. These two freshmen, then only 26 years old, challenged the forefront problems of nuclear physics, and attended the spring meeting of the Japan Physico-Mathematical Society, held in Sendai in 1933. 

\begin{figure}[tbh]
\begin{center}
\includegraphics[width=8cm]{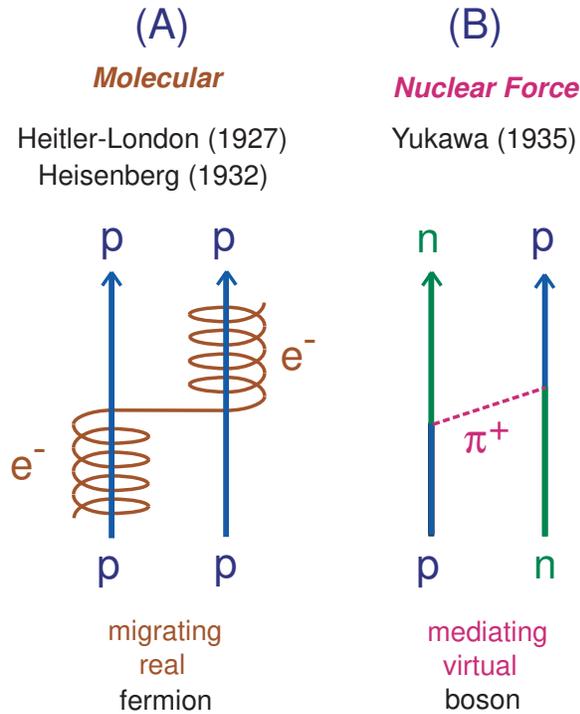}
\end{center}
\caption{\label{fig:HLH} (Left) Heitler-London-Heisenberg scheme for the nuclear force. Diagram drawn  by K. Nishijima. (Right) Yukawa's meson exchange interaction for the nuclear force.} 
\end{figure}

At that time Heisenberg was stuck to the idea of molecular binding applied to the nuclear force in terms of ``Platzwechsel" {\it a la} Heitler and London \cite{Heitler:27}, who explained for the first time the H-H bonding in the hydrogen molecule quantum mechanically in 1927. The Heitler-London-Heisenberg mechanism can be expressed by a diagram, as shown in Fig.~\ref{fig:HLH} (Left). An electron migrates between the two protons, and a strong bonding force emerges as a quantum mechanical effect. Heisenberg's attempt  was to combine the proton and the neutron in a similar fashon.

Yukawa took up this problem seriously, and presented his struggling with this problem at this Sendai  meeting. The title of his talk was ``a consideration of the problem of nuclear electrons". At this meeting Yoshio Nishina, who is called Father of modern physics in Japan, stimulated Yukawa and Tomonaga toward the then developing quantum physics,  suggested that Yukawa should consider a boson instead of a fermion to avoid one of the difficulties. The term ``Bose electron" was used at that time. It took some time until Yukawa reached his revolutional idea of a mediating virtual meson in the fall of 1934. 

At the same meeting Tomonaga presented his work on the deuteron binding energy and the proton-neutron reaction. At that time Tomonaga was a resident physicist in Nishina's laboratory at RIKEN, and was working on a     
theoretical explanation of the newly obtained experimental data on the interaction between a proton and  a neutron, employing various interaction forms. It is striking that he chose the following form for a short-range interaction: 
\begin{equation}
J(r) = A\, \frac{{\rm exp} (- \lambda r)}{r}.\label{eq:Yukawa-form}
\end{equation}
This was nothing but what would be later called the Yukawa interaction.

Here, a great interplay emerged. After this meeting Tomonaga wrote a rather long letter to Yukawa \cite{Tomonaga:33}, in which he explained his results in more detail. Figure~\ref{fig:Tomonaga-letter} is a copy of his letter, in which he showed the above interaction form, and told Yukawa about the value of the range parameter  $\lambda$ he obtained from fitting the experimental values with this interaction form,
\begin{equation}
\lambda = 7 \times 10^{12}~/{\rm cm.}
\end{equation}

\begin{figure}[tbh]
\begin{center}
\includegraphics[width=8cm]{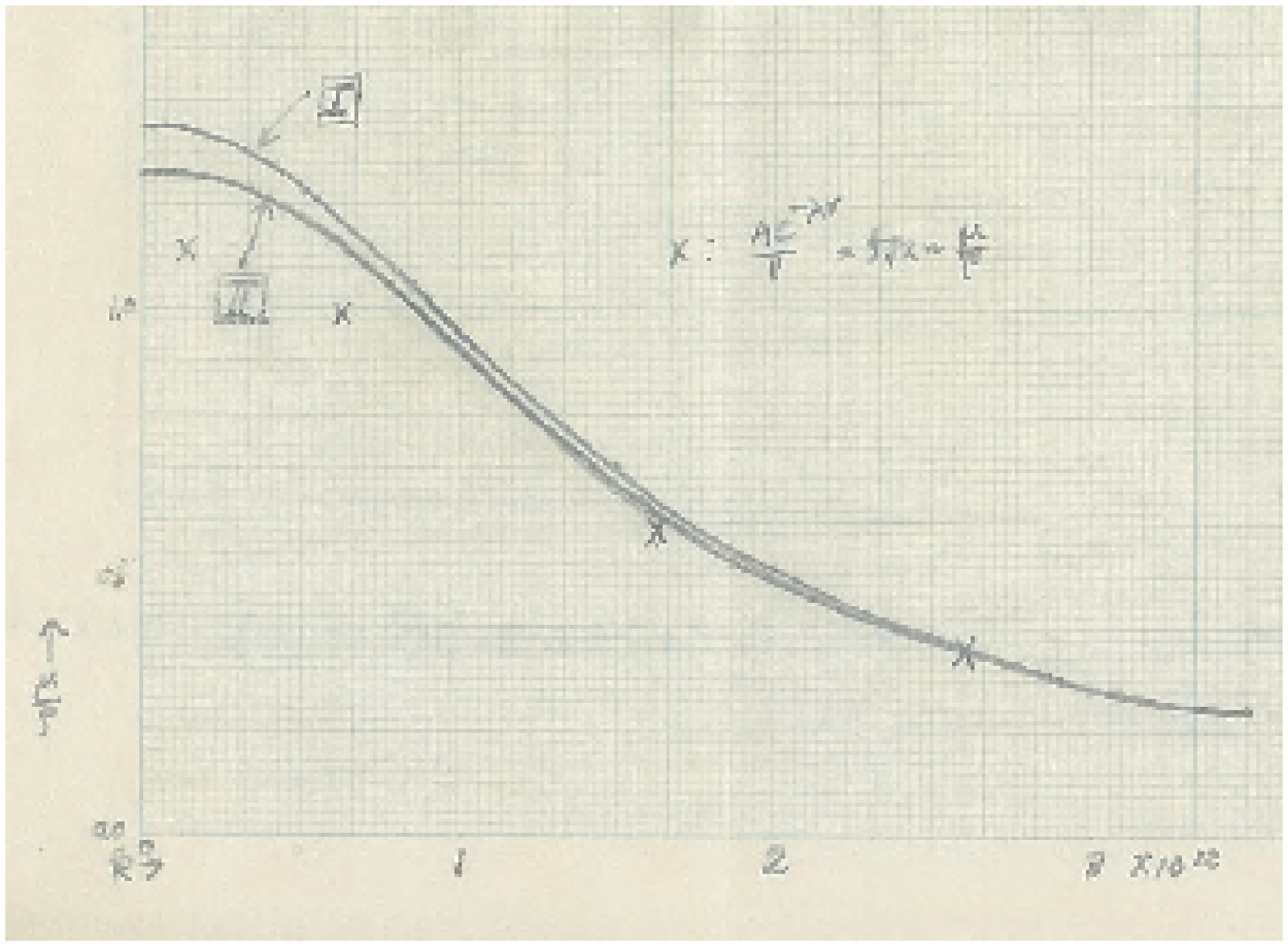}
\includegraphics[width=4.5cm]{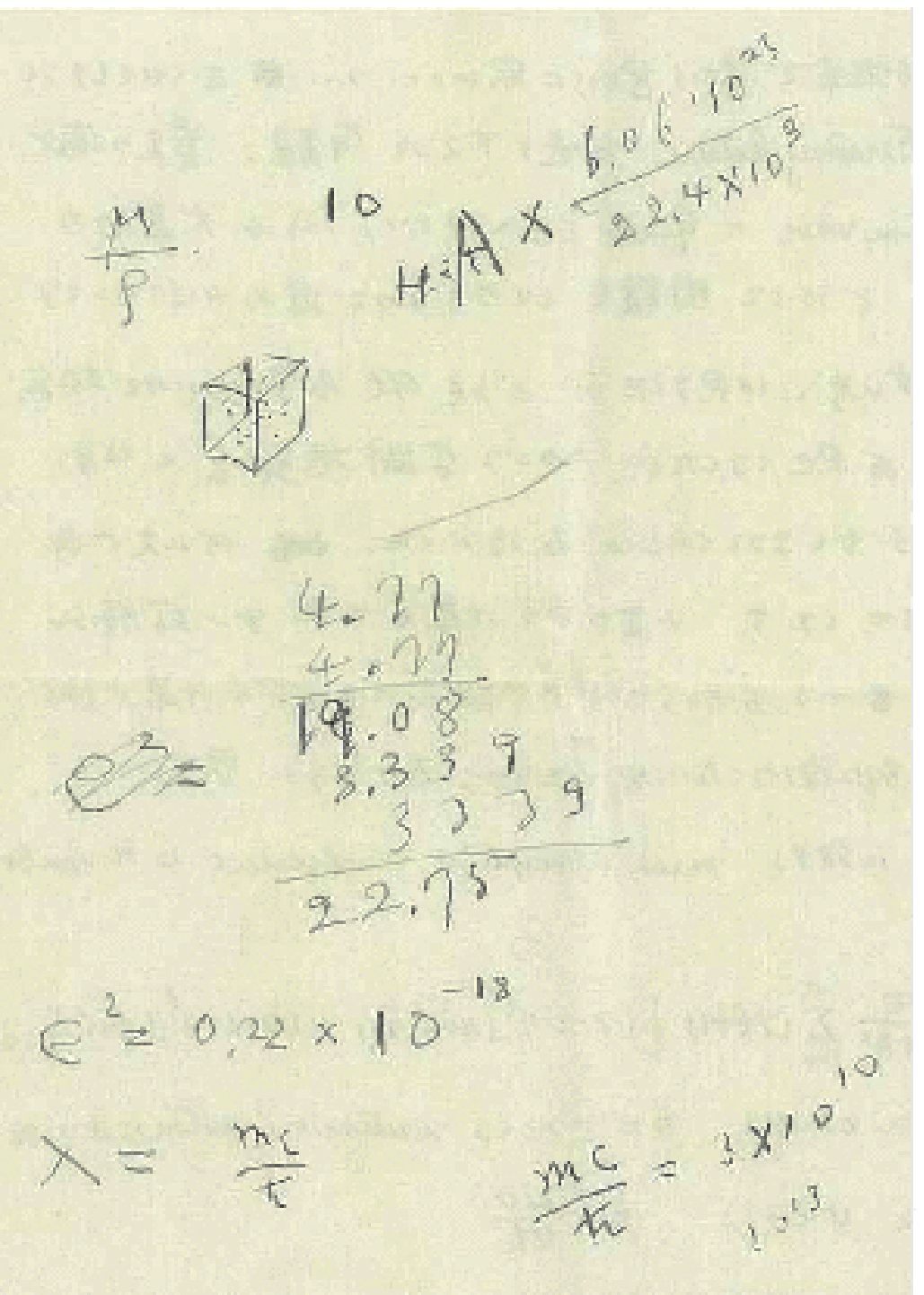}
\end{center}
\caption{(Left) Tomonaga's hand-drawn plot of the p-n reaction data with his theoretical fitting using the ``Yukawa interaction."  (Right) Yukawa's note on the back of Tomonaga's letter. 
From Ref. \cite{Tomonaga:33}.} 
\end{figure}\label{fig:Tomonaga-letter}

 It is extremely interesting that Yukawa jotted some notes on the back of this letter (see Fig.~\ref{fig:Tomonaga-letter}, right panel), such as  
\begin{equation}
\lambda_{Compton} = \frac{mc}{\hbar} \sim 3 \times 10^{10}~{\rm cm}^{-1}.
\end{equation} 
It is very interesting to speculate about what Yukawa was thinking when he made this hand-written estimate of the electron Compton wavelength. If we divide Tomonaga's value by this value, we would obtain the value 230 !! 

We can thus imagine that this letter must have had a profound influence on  Yukawa, who was in the midst of struggling with the problem of the nuclear force in 1933, but had not yet formulated the idea of  the Yukawa interaction, in which the range parameter is related to the mass of the mediating particle. 

Tomonaga's work on the range of the p-n interaction was later mentioned in a footnote of Yukawa's first paper \cite{Yukawa:35} as\\

\begin{quotation}
{\it These calculations were made previously, according to the theory of Heisenberg, by Mr. Tomonaga, to whom the writer owes much....}\\
\end{quotation}

\noindent
On the other hand, Tomonaga published this work only in 1936 \cite{Tomonaga:36}, 3 years after his letter to Yukawa. A similar work on the proton-neutron binding by Bethe and Peierls \cite{Bethe:35} appeared in the literature in 1935.

\section{Yukawa Overcame Heisenberg}

 At that time, in the spring of 1933, neither Yukawa nor Tomonaga seemed to recognize the deep meaning of the interaction formula (\ref{eq:Yukawa-form}) and, in particular, the implication of the range parameter, $\lambda$, which Tomonaga deduced from experimental data. 
 Heisenberg tried to explain the nuclear force in terms of the Heitler-London scheme, but was not successful. Yukawa as of 1933 also struggled with this problem. Many difficult problems existed: how can a migrating real particle exist in nuclei? Is it a Bose electron? How can the interaction be short-ranged? How can proton + electron be a neutron?

Finally, in the fall of 1934 Yukawa arrived at the concept of a ``virtual mediating particle" behind the strong  nuclear force. Its diagram is shown in Fig.~\ref{fig:HLH} (Right).

The famous equation
\begin{equation}
 [\Delta - \frac{1}{c^2} \, \frac{\partial ^2}{\partial t^2} - \lambda^2] \, U = 0,
\end{equation} 
given in the first Yukawa paper, published in 1935 \cite{Yukawa:35}, represents a force field $U$ with a parameter $\lambda$. Here, Yukawa took the form (\ref{eq:Yukawa-form})
for the short-range interaction of the nuclear force. Yukawa's great discovery was to relate this parameter,  which expresses the inverse of the force range, to the mass of a mediating particle, $U$:
\begin{equation}
\lambda = \frac{m c}{\hbar}.
\end{equation} 
The mass of $U$ was expected to be 200-times larger than the electron mass. Eventually, the Yukawa mesons were discovered. Thus, the Yukawa theory became the fundamental concept in modern physics since 1935.

In contrast, the old idea of Heisenberg based on the Heitler-London scheme was completely abandoned and forgotten in nuclear physics, but we will revisit  this problem later.

\section{Discovery of the ``Mesotron"}

\begin{figure}[tbh]
\centering
\includegraphics[width=6.5cm]{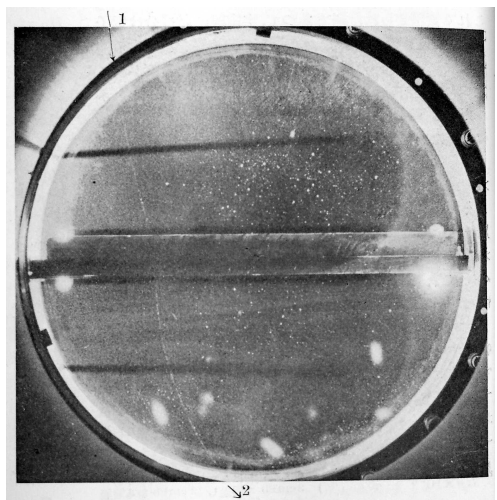}
\caption{Photograph of the cloud chamber track of a cosmic ray event taken by Nishina, Takeuchi and Ichimiya of RIKEN. From Nishina in ``Kagaku" \cite{Kagaku}.  }\label{fig:muon-track}
\end{figure}

Another story I would like to convey is that of the discovery of muons. Nishina constructed a large cloud chamber with a very strong and homogeneous magnetic field to measure cosmic rays. Around 1936-37,  there were four experimental groups in the world with the primary purpose of examining the validity of the Bethe-Heitler formula, which had just been derived. Neddermeyer and Anderson \cite{Nedd} were the first to report that there are some particles that do not obey this theory. Such particles were believed to be neither the proton nor the electron (positron), presumably having a mass between that of the proton and the electron. In the same year, similar findings were reported by other groups \cite{Jean,Nishina-muon,Street}. Among them, two groups succeeded in determining the mass of such intermediate particles. A paper of  Nishina, Takeuchi and Ichimiya \cite{Nishina-muon}, reporting a value of $m_X/m_e = 180 \pm 20$, the most precise value at that time, was received by Physical Review on August 28, 1937, and was published on December 1. Interestingly, a paper of Steet and Stevenson \cite{Street}, reporting $m_X/m_e = 130 \pm 30$, was received on October 6, 1937, more than one month later than Nishina's paper, but was published on November 1, one month earlier. This situation resulted from the fact that shipping 
 of the galley proof back and forth between New York and Tokyo took nearly 40 days. Figure \ref{fig:muon-track} shows a cloud-chamber picture of Nishina' group, which was printed in a Japanese science journal,  ``Kagaku",  in September 1937. 

Thus, it is fair to say that the two experiments \cite{Nishina-muon,Street} were nearly of the same quality and significance. Nevertheless, the experiment of Nishina's group has hardly been recognized in the world. It is a pity that even Japanese physicists are not aware of this great achievement. \\

\section{Heitler-London-Heisenberg scheme revisited}

\begin{figure}[htb]
\centering
\includegraphics[width=12cm]{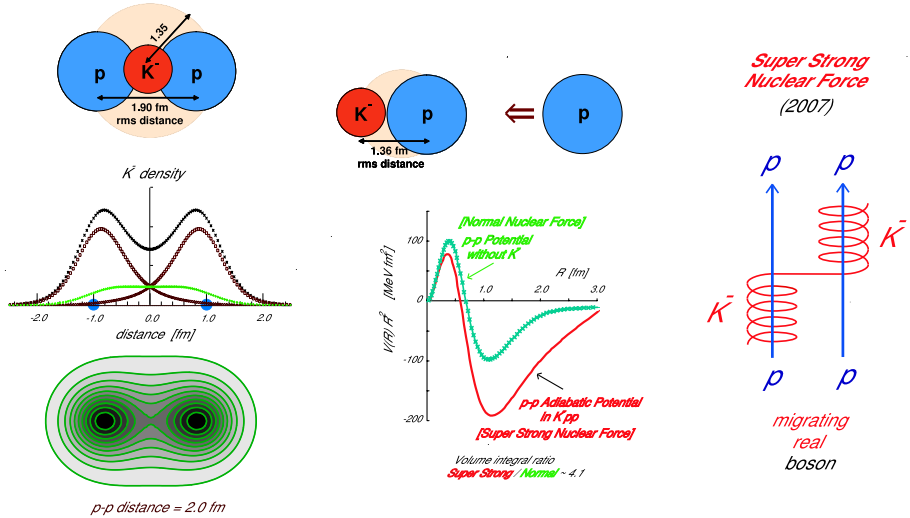}
\vspace{0cm}
\caption{\label{fig:SSNF} 
(Right) Molecular structure of $K^-pp$. The projected density distributions of $K^-$ in $K^-pp$ with a fixed $p-p$ distance (= 2.0 fm) and the corresponding $K^-$ contour distribution are shown, 
(Center) Adiabatic potential ($V(R) \, R^2$), when a proton approaches a bound $K^-p$ ``atom" ($\Lambda^*$), as a function of the distance between $p$ and $p$. For a comparison the Tamagaki potential for the normal $V_{NN}$ interaction is shown.
(Right) Super strong nuclear force  caused by a migrating real boson, $\bar{K}$, expressed similar to the Heitler-London-Heisenberg covalency. }
\end{figure}

My historical talk may end at this point, but I would like to show two more slides to explain that the Heitler-London-Heisenberg scheme may have profound meaning, 80 years after the Heitler-London paper.
Very recently Akaishi and myself predicted the presence of a new nuclear cluster, $K^- p p$ based on the empirically deduced $\bar{K} N$ interaction \cite{Akaishi:02,Yamazaki:02}. In studying its structure from an unconstrained three-body calculation, we found a dynamically organized molecular structure with $K^-p$ as an atomic constituent \cite{Yamazaki:07a,Yamazaki:07b}, as shown in Fig.~\ref{fig:SSNF} (Left). This situation brings a new type of force, called the ``super strong nuclear force", which can be compared with the ordinary nuclear force, as visualized in Fig.~\ref{fig:SSNF} (Middle).

A distinct difference is seen in the depth of the attractive potential and the range of the potential. The volume ratio amounts to about 4. This super strong nuclear force, thus introduced, might be the cause for a cold dense nuclear system and kaon condensation \cite{Kaplan:86,Brown:94}. Thus, we have shown that the HLH scheme is revived in a new form of the nuclear force, and its experimental verification in the future would be extremely interesting. \\

The author would like to thank Prof. Y. Akaishi for collaborative work. This work is supported by a Grant-in-Aid of Monbukagakusho of Japan.

\end{document}